\documentclass{aa}
\usepackage{txfonts}
\usepackage{graphicx}
\usepackage{natbib}
\bibpunct{(}{)}{;}{a}{}{,}
       
\begin{document}

\def\vx{\vec{x}}
\def\vr{\vec{r}}
\def\vxp{\vec{x}_\perp}
\def\vrp{\vec{r}_\perp}
\def\vk{\vec{k}}
\def\vkp{\vec{k}_\perp}
 
\def\RM{{\rm RM}}
\def\obs{{\rm obs}}
\def\rf{{\rm ref}}
\def\expect{{\rm exp}}
\def\eps{\varepsilon}

\title{Measuring the Cluster Magnetic Field Power Spectra from Faraday
Rotation Maps of\\Abell 400, Abell 2634 and Hydra A}
\titlerunning{Measuring Cluster Magnetic Field Power Spectra}
\author{Corina Vogt \and Torsten A. En{\ss}lin}
\authorrunning{Corina Vogt \and Torsten A. En{\ss}lin}
\institute{Max-Planck-Institut f\"{u}r
Astrophysik, Karl-Schwarzschild-Str.1, Postfach 1317, 85741 Garching,
Germany} 

\date{Submitted 06 June 2003/ Accepted}

\abstract{
We apply a novel technique of Faraday Rotation measure (RM) map
analysis to three galaxy clusters, Abell 400, Abell 2634 and Hydra A,
in order to estimate cluster magnetic field strengths, length scales
and power spectra. This analysis - essentially a correlation analysis
- is based on the assumption that the magnetic fields are
statistically isotropically distributed across the Faraday screen. We
investigate the difficulties involved in the application of the
analysis to observational data. We derive magnetic power spectra for
three clusters, i.e. Abell 400, Abell 2634 and Hydra A, and discuss
influences on their shapes caused by the observational nature of the
data such as limited source size and resolution. We successfully apply
various tests to validate our assumptions. We show that magnetic
fluctuations are probed on length scales ranging over at least one
order of magnitude. Using this range for the determination of magnetic
field strength of the central cluster gas yields 3 $\mu$G in Abell
2634, 6 $\mu$G in Abell 400 and 12 $\mu$G in Hydra A as conservative
estimates. The magnetic field autocorrelation length $\lambda_B$ was
determined to be 4.9 kpc for Abell 2634, 3.6 kpc for Abell 400 and 0.9
kpc for Hydra A. We show that the RM autocorrelation length
$\lambda_{\RM}$ is larger than the magnetic field autocorrelation
length $\lambda_B$ - for the three clusters studied, we found
$\lambda_{\RM} \simeq 2...4\lambda_B$ - and thus, they are not equal as
often assumed in the literature. Furthermore, we investigate in a
response analysis if it is possible to determine spectral slopes of
the power spectra. We find that integrated numbers can be reliably
determined from this analysis but differential parameters such as
spectral slopes have to be treated differently. However, our response
analysis results in spectral slopes of the power spectra of spectral
indices $\alpha$ = 1.6 to 2.0 suggesting that Kolmogorov spectra
($\alpha = 5/3$) are possible but flatter spectral slopes than $\alpha
= 1.3$ can be excluded. 

 }

\maketitle

\section{Introduction\label{sec:intro}}
Clusters of galaxies are inhabited by magnetised plasma. The detection
of diffuse synchrotron emission observed as cluster radio halos and
radio relics is but one piece of evidence for the existence of cluster
wide magnetic fields \citep[for a review,
see][]{2002ARA&A..40..319C}. Still, their strengths, their structures
and their origin remains to be unveiled.

Although the cluster wide magnetic fields are not believed to
influence the dynamics directly in most clusters because the magnetic
pressure is estimated to be too small compared to the thermal
pressure, they are still important for the understanding of the
physics of the intra-cluster gas. Indirectly they are likely an
important cluster gas ingredient, e.g. cluster magnetic fields
suppress heat conduction as observed in cluster cold fronts
\citep[e.g.][]{2000ApJ...541..542M, 2001ApJ...549L..47V}.
 
There are different methods to gain information about magnetic
fields. One is to use the synchrotron radiation observed as radio
halos or relics to estimate the field strength. Energy equipartition
arguments for these clusters lead to strengths of about $\sim$ 0.1...2
$\mu G$ \citep[e.g.][]{1999dtrp.conf....3F,
2001MNRAS.320..365B}. Another possibility is the comparison of their
synchrotron radiation fluxes with inverse Compton X-ray emission
leading to lower limits to the field strengths of about 0.2...1 $\mu$G
\citep[e.g.][]{1979MNRAS.188...25H, 1987ApJ...320..139R,
1998A&A...330...90E, 1998tx19.confE.513F, 1998MNRAS.296L..23B}.

A different method to gain knowledge about magnetic fields is the
observation of the Faraday rotation effect which arises whenever
polarised radio emission passes through a magnetised medium, causing
its plane of polarisation to rotate if the magnetic field component
parallel to the direction of the light propagation is nonzero. If the
Faraday active medium is external to the source of the polarised radio
emission one expects the rotation to be proportional to the squared
wavelength. The proportionality constant is called the Faraday rotation
measure (RM) which can be evaluated in terms of the line of sight
integral over the product of the electron density and the magnetic
field component along the line of sight.

Recently, \citet{2003ApJ...588..143R} claimed to have evidence for the
RM producing fields being local to the polarised radio source as
suggested by \citet{1990ApJ...357..373B}. However, this conclusion was
based on a biased statistic without performing a proper
null-experiment \citep{2003ApJ...Ensslin}. Furthermore, the assumption
that the Faraday rotation is produced in the intra-cluster gas is also
supported by independent evidence. We mentioned already some of this
evidence, the cluster wide diffuse radio emission observed as radio
halos and relics. Further evidence is the observation of the
Laing-Garrington effect. It manifests itself by an asymmetric Faraday
depolarisation of double radio lobes located within galaxy clusters
\citep{1988Natur.331..147G, 1988Natur.331..149L}. This asymmetry is
caused by different path lengths for radio emission through the
Faraday active medium between source and observer as suggested by
\citet{1991MNRAS.250..198G}. This effect could also  be explained by
an asymmetric mixing layer being thin close to the head side of a FR
II while getting thicker on the back-flow side of it. However, this
explanation fails to give a convincing explanation for the large RM
and depolarisation asymmetry observed for the FR I source Hydra A
\citep{1993ApJ...416..554T}.

Further independent evidence for cluster wide magnetic fields is
provided by a recent statistical RM investigation of point sources
performed by \citet{2001ApJ...547L.111C}. It revealed larger RM values
for sources observed through the intra-cluster gas in comparison to a
control sample of sources where no intra-cluster gas was located
between source and observer. This suggests that the enhancement of the RM
towards the cluster centre results most probably from the magnetised
cluster gas.

Thus, we believe that the analysis of Faraday rotation maps of
extended polarised radio sources located behind or embedded in galaxy
clusters allows us to understand the strength and to get some hints
about the structure of cluster magnetic fields.

A naive analysis of such maps is to estimate a minimum field strength
by assuming a constant magnetic field throughout the cluster along the
line of sight. However, close inspection of RM maps of extended
extragalactic radio sources reveals a patchy structure on small scales
(5 - 20 kpc) which indicates that the fields are tangled. An improved
analysis of the RM maps assumes the Faraday screen consists of cells
which have a constant size and a constant magnetic field strength but
the field direction varies randomly from cell to cell. The mean RM
which would be produced by such a screen builds up in a random walk
and thus has a zero mean but a non vanishing dispersion. This
dispersion will be proportional to the square root of the number of
cells along the line of sight, the cell size, the electron density
profile and the magnetic field strength as shown by
\citet{1996clfu.conf..271F} and \citet{1995A&A...302..680F}. Using
these proportionalities magnetic field strengths of the order of a few
$\mu$G for non-cooling flow clusters,  e.g. Coma
\citep{1995A&A...302..680F}, A119 \citep{1999A&A...344..472F}, 3C129
\citep{2001MNRAS.326....2T}, A514 \citep{2001A&A...379..807G}, A400 \&
A2634 \citep{2002ApJ...567..202E}, and of the order of a few 10 $\mu$G
for cooling flow clusters, e.g. Cygnus A \citep{1987ApJ...316..611D},
3C295 \citep{1991AJ....101.1623P}, A1795 \citep{1993AJ....105..778G},
Hydra A \citep{1993ApJ...416..554T}, were derived.
 
However, there are some drawbacks of the cell model. First of all, the
divergence of the magnetic field in such a Faraday screen would be
nonzero which contradicts Maxwell's equations. Secondly, a spectrum of
scales for the cell sizes is more likely than a single scale.

Another important issue in previous analyses is the assumption that
the RM ordering scale read off RM maps is equivalent to the magnetic
field's characteristic length scale, the autocorrelation length
$\lambda_B$. This might have lead to underestimations of field
strengths in the past since the equation often used for the derivation
of the magnetic field strength is
\begin{equation}
\label{eq:bzlb}
\langle B^2 _z \rangle = \frac {\langle \RM^2 \rangle}{a_0 ^2\, n_{e}
^2\, L\, \lambda _z} ,
\end{equation}
where $\langle \RM ^2 \rangle$ is the RM dispersion, $L$ is the depth
of the Faraday screen, $a_0\,=\,e^3/(2 \pi m_e^2c^4)$, $n_e$ is the
electron density and $\lambda _z$ is a characteristic length scale of
the fields.  At this point it becomes clear that the definition of the
characteristic length scale $\lambda _z$ is crucial for the derivation
of magnetic field strength. The assumption of a constant magnetic
field throughout the cluster leads to a definition of $\lambda _z = L$
and thus, $\langle B^2 _z \rangle \propto L^{-2}$. The characteristic
length scale for the cell model would be the size of each cell
$\lambda _z = l_{cell}$. Another definition one could think of is the
RM ordering scale, $\lambda_z = \lambda_{\RM}$. However, the correct length
scale is the autocorrelation length $\lambda _z$ of the
$z$-component of the magnetic field measured along the line of sight.

A purely statistical approach which incorporates the vanishing divergence
of the magnetic field was recently developed by
\citet{2003A&A...401..835E}. This approach relies on the assumption
that the fields are statistically isotropically distributed in Faraday
screens such that the $z$-component is representative for all
components. Starting from this assumption, a relation between the
observationally accessible RM autocorrelation function and the magnetic
autocorrelation tensor is established in real and in Fourier space
such that one gains access to the power spectrum of the magnetic field
inhabited by the Faraday screen.

\citet{2003A&A...401..835E} demonstrated that in the case of isotropy
the magnetic autocorrelation length $\lambda _z$ in Eq.(\ref{eq:bzlb})
is given by $\lambda _z = 3/2\,\lambda_B$, where $\lambda_B$ is the
3-dimensional magnetic autocorrelation length. Therefore,
Eq. (\ref{eq:bzlb}) becomes
\begin{equation}
\langle B^2 \rangle = 3\, \langle B^2 _z \rangle = \frac {2\,\langle
\RM^2 \rangle}{a_0 ^2\, n_{e} ^2\, L\, \lambda _B}.
\end{equation}
The length scale $\lambda_B$ can be estimated from the measured RM power
spectrum.

Here we apply this statistical analysis to observational data by
reanalysing the Faraday rotation measure maps of three extended
extragalactic radio sources: Hydra A \citep{1993ApJ...416..554T}, 3C75
and 3C465 \citep{2002ApJ...567..202E} which were kindly provided by
Greg Taylor, Frazer Owen, and Jean Eilek, respectively. 

In section \ref{sec:method}, we outline shortly the idea of our
approach, the real and the Fourier space analysis and what
possibilities are given to test if the assumptions made are
reasonable. Furthermore, we state the equations which were used in the
analysis while their derivation and more detailed discussion can be
found in \citet{2003A&A...401..835E}. In  section
\ref{sec:application}, the application of these approaches to the
observational data and the difficulties involved are
described. Emphasis is given to a critical discussion of the strengths
and limits of the approach. The results are  presented and discussed
in section \ref{sec:discussion}. Finally, the conclusions are  drawn in the
last section \ref{sec:conclusion}.

Throughout the rest of the paper we assume a Hubble constant of
H$_{0} =  70$ km s$^{-1}$ Mpc$^{-1}$, $\Omega_{m} = 0.3$ and
$\Omega_{\Lambda} = 0.7$ in a flat universe. All equations follow the
notation of \citet{2003A&A...401..835E}.

\section{Method\label{sec:method}}
\subsection{Idea \label{sec:model}}

We assume that the magnetic fields in galaxy clusters are
isotropically distributed throughout the Faraday screen. If one
samples these field distributions over a sufficiently large volume
they can be treated as statistically homogeneous and statistically
isotropic. This means that the statistical average over any field
quantity will not be influenced by the geometry or the exact location
of the volume sampled.

One of the magnetic field quantities which is of interest here is its
autocorrelation function since this function can be used to calculate
its correlation length and field strength. Furthermore, the
autocorrelation function is equivalent to the Fourier transform of the
power spectrum as stated by the Wiener-Khinchin Theorem (WKT).

Since the magnetic field is a vector quantity, an autocorrelation
tensor has to be considered rather than a function. The diagonal
elements of this tensor are all equal in the case of magnetic isotropy
which we assume as stated above. On the other hand, its off-diagonal
terms consist of two numbers describing the symmetric and
non-symmetric contribution. The divergence-freeness of magnetic fields
($\vec{\nabla \cdot B} = 0$) couples the symmetric components of the
tensor which means that the knowledge of a diagonal element ensures
complete knowledge of the symmetric part of the tensor. Introducing
now the \emph {scalar magnetic autocorrelation function} $w(r)$  as
the trace of the autocorrelation tensor $w(r) = \langle \vec{B}(\vx)
\cdot \vec{B}(\vx + \vr) \rangle _{x}$, field quantities can be
derived such as the magnetic autocorrelation length and
average magnetic field energies using this function.

The non-symmetric part of the tensor represents the magnetic helicity.
These off-diagonal terms of the tensor do not enter in the definition
of the scalar magnetic autocorrelation function $w(r)$ and thus, do
not have any effect on the quantities of interest listed above.

The observable considered in our approach is the Faraday rotation
which is related to the projected magnetic field distribution along
the line of sight. It is possible to establish a relation between the
autocorrelation function of the RM map and the autocorrelation
function of the magnetic field as discussed in more detail in the next
sections.

As stated above, the knowledge of the autocorrelation function gives
access to the power spectra using the WKT. However, any phase
information is lost by measuring power spectra. Therefore, it is
possible that different realisations of magnetic field structure have
the same power spectrum. This implies that not all information which might be
contained in the RM maps are accessible by the analysis
presented. Correlation function of higher order statistic are
needed to access these information.

In the discussion so far, we ignored the fact that we are using
observational data and that the RM is also dependent on electron
density and magnetic energy density on global scales which makes the
introduction of a window function necessary. This window function can
be interpreted as the sampling volume. It accounts for the limited
size of the radio source and the global properties of the cluster gas
such as the electron density and the average magnetic energy density
profiles. A virtually statistical homogeneous magnetic field can be
thought to be observed through this function.

The window function will be zero for points outside the volume which
is situated in front of the radio source. It will scale with the
electron density and the global magnetic field energy distribution
within the Faraday screen between radio source and
observer. Furthermore if wanted the window function can contain noise
reducing data weighting schemes.

The overall effect of any finite window in real space is to smear out
power in Fourier space. Furthermore, any analysis employing a window
function will be sensitive to magnetic energies on scales smaller than
the window size and insensitive to scales larger than the window
size. Therefore, understanding the influence of the window function is
crucial for the evaluation of the results obtained by applying of our
analysis to observational data. Possibilities to assess its influence
are given in section \ref{sec:testmodel}.

\subsection{Real Space Analysis \label{sec:spatana}}
For a line of sight parallel to the z-axis and displaced by $\vxp$
from it, the Faraday rotation measure arising from polarised emission
passing from the source at $z_{\rm s}(\vxp)$ through a magnetised
medium to the observer located at infinity is expressed by
\begin{equation}
\label{eq:RMbasic}
\RM(\vxp)=a_0 \int_{z_{\rm s}(\vxp)}^{\infty}
\!\!\!\!\!\!dz\,n_e(\vx)\,B_z(\vx),
\end{equation}
where $a_0\,=\,e^3/(2 \pi m_e^2c^4)$, $\vx\,=\,(\vxp,z)$, $n_e(\vx)$
is the electron density and $B_z(\vx)$ is the magnetic field component
parallel to the line of sight.

The autocorrelation function of the Faraday rotation map can be
defined as
\begin{equation}
C_{\RM}(\vrp) = \langle \RM(\vxp)\RM(\vxp+\vrp) \rangle _{\vxp},
\end{equation}
where the brackets mean the map average with respect to
$\vxp$. However, an observed RM map is limited in size which leads to
noise in the calculation of the correlation function on large scales
since less pixel  pairs contribute to the correlation function for
larger pixel separations. For suppressing this under-sampling,
the observable correlation function was chosen to be
\begin{equation}
\label{eq:crm_real}
C^{\obs}_{\RM} (r_{\perp})=\frac{1}{A_{\Omega}} \int
dx^{2}_{\perp}\,\RM(\vxp)\RM(\vxp+\vrp), 
\end{equation}
where the area $A_{\Omega}$ of the region $\Omega$ for which RM's are
actually measured is used as normalisation assuming that $\RM(\vxp) =
0$ for $\vxp \notin \Omega$. Furthermore, it is required that the mean
RM is zero since statistically isotropic divergence free fields have a
mean RM of zero to high accuracy. The non vanishing mean RM in the
observational data stems from foregrounds (e.g. the galaxy) in which
we are not interested or has its origin in large scale fields to which
the approach is insensitive by construction in order to suppress
statistically under-sampled length scales.

As stated in section \ref{sec:model}, the virtually homogeneous
magnetic field component can be thought to be observed through a
window function $f(\vx)$ which describes the sampling volume. One
would chose a typical position in the cluster $\vx_{\rf}$ (e.g. its
centre) and define $n_{e0} = n_e(\vx_{\rf})$ and $\vec{B} _{0} =
\langle \vec{B} ^2 (\vx_{\rf}) \rangle ^{1 / 2}$. The window function
can then be expressed as
\begin{equation}
f(\vx) = \mathbf{1}_{\{\vxp \in \Omega\} }\,\mathbf{1}_{\{z \geq
z_{\rm s}(\vxp)\}} \,h(\vxp) \,g(\vx) \,n_e(\vx)/n_{e0},
\end{equation}
where $\mathbf{1}_{\{condition\}}$ is equal to one if the condition is
true and zero if not. The dimensionless average magnetic field profile
$g(\vx) = \langle \vec{B} ^2 (\vx) \rangle ^{1 / 2} / \vec{B} _{0}$ is
assumed to scale with the density profile such that $g(\vx) =
(n_e(\vx)/n_{e0})^{\alpha_{B}}$. Reasonable values for the exponent
$\alpha_{B}$ range between 0.5 and 1 as suggested by
\citet{2001A&A...378..777D}. The function $h(\vx)$ allows
to introduce a suitable data weighting scheme and can be chosen to be
unity if no data weighting is required.

As mentioned above, the magnetic autocorrelation function $w(r)$ can
be defined as the trace of the magnetic field autocorrelation tensor
$M_{ij} = \langle B_i (\vx)\,B_j(\vx+\vr) \rangle$ assuming isotropy
$w(\vr) = \langle \vec{B}(\vx) \cdot \vec{B}(\vx+\vr) \rangle = \sum
_i M_{ii}$, \citep[e.g.][]{1999PhRvL..83.2957S}. However, the Faraday
rotation is also related to the magnetic field via
Eq.(\ref{eq:RMbasic}). Therefore any RM correlation contains also
information on the autocorrelation of the magnetic field component
along the line of sight 
\begin{eqnarray}
\label{eq:mcomp}
C_{\perp}(r_{\perp}) &=& \int ^{\infty} _{-\infty} dr_{z} M_{zz}(\vr)\\
\label{eq:crwr_real}
&=& \frac {1}{2} \int ^{\infty} _{-\infty} dr_{z} w(\sqrt{r^2 _{\perp}
+ r^2 _{z}}), 
\end{eqnarray}
where $\vr = (\vrp, r_z)$. Taking the effect of the window function
into account, we introduced $C_{\perp}(r_{\perp}) = \langle C^{\obs}
_{\RM} (\vrp) \rangle /a_1$, where $a_1 = a_0 ^2 n_{e0} ^2 L$. The
parameter $L$ can be interpreted as the characteristic depth of the
Faraday screen and can be expressed as $L = V_{[f]}/A_{\Omega}$, where
$V_{[f]}$ is the probed effective volume $\int dx^3 \, f^2(\vx)$.

Using Eq. (\ref{eq:crwr_real}), one is now able to establish a
relationship between the RM autocorrelation function
$C_{\perp}(r_{\perp})$ and the magnetic autocorrelation function
$w(r)$. This relationship can be expressed as
\begin{eqnarray}
\label{eq:wr1_real}
w(r) &=& - \frac {2}{\pi r} \frac {d}{dr} \int ^{\infty}_{r} dy
\frac{y\,C_{\perp}(y)}{\sqrt{y^2-r^2}}\\
\label{eq:wr2_real}
 &=&-\frac{2}{\pi} \int ^{\infty} _{r} dy \frac {C'(y)}{\sqrt{y^2-r^2}},
\end{eqnarray}
where for the deprojection of Eq. (\ref{eq:crwr_real}) an Abel
integral equation was used assuming $w(r)$ stays bounded for $r \to
\infty$. 

These two equations enable us to calculate not only the magnetic
autocorrelation function but also to obtain directly the average
magnetic energy density $\langle \varepsilon _{B} \rangle$ by using
the relation $\langle \varepsilon_B \rangle = w(0) / 8\pi$ and thus,
we are able to determine the magnetic field strength by calculating
the value of the magnetic autocorrelation function at point $r = 0$:
$w(r = 0) = \langle B^2 \rangle$.

The RM autocorrelation length $\lambda_{\RM}$ and the magnetic
autocorrelation length $\lambda_{B}$ can be calculated by
integrating the correlation functions:
\begin{eqnarray}
\lambda_{\RM} &=& \int ^{\infty} _{-\infty} dr \frac{\langle
\RM(\vx+r\vec{e_{\perp}})\,\RM(x) \rangle
_{\vx,\vec{e_{\perp}}}}{\langle \RM ^2 (\vx) \rangle} \\ &=&\int
^{\infty} _{-\infty} dr_{\perp} \frac
{C_{\perp}(r_{\perp})}{C_{\perp}(0)} = \pi \frac {\int ^{\infty}
_{-\infty} dr \,r\, w(r)}{\int ^{\infty} _{-\infty} dr\, w(r)}
\label{eq:lrm_real}\\
\lambda_{B} &=& \int ^{\infty} _{-\infty} dr \frac{\langle
B(\vx+r\vec{e})\, \vec{\cdot}\, B(\vx) \rangle _{\vx,\vec{e}}}{\langle
B^2 (\vx) \rangle} \\ &=& \int ^{\infty} _{-\infty} dr \frac
{w(r)}{w(0)} = 2 \frac {C_{\perp}(0)}{w(0)},
\label{eq:lb_real}
\end{eqnarray}
where $\vec{e}$ and $\vec{e}_{\perp}$ are 3- and 2-dimensional unit
vectors, respectively, over which we are averaging.

\subsection{Fourier Space Analysis \label{sec:fourana}}
In Fourier space a similar formalism can be developed assuming
statistical isotropy and relying on the $\vec{\nabla \cdot B} = 0$
condition. After performing a 2-dim Fourier transform\footnote
{
defining the Fourier transformation for a $n$-dimensional function
$F(\vx)$ such that:

\begin{math}
\hat{F}(k)\,=\,\int d^n x\,F(\vx)\,e^{i\vk \cdot \vx};
F(\vx)\,=\,\frac{1}{(2\pi)^n} \int d^n k\,\hat{F}(\vk)\,e^{-i\vk \cdot
\vx}.
\end{math}
} 
of the RM map, one can calculate the RM autocorrelation in Fourier
space:
\begin{equation}
\label{eq:crm_four}
\hat{C}_{\perp}(k_{\perp})= \frac {\langle \,|\, \hat{\RM} (k_{\perp})
|^2 \,\rangle} {a_1 A_{\Omega}},
\end{equation} 
implied by the WKT which connects the 2-dimensional Fourier
transformed RM maps with the RM autocorrelation function.

In Fourier space Eq.(\ref{eq:crwr_real}), the relationship between the
RM and the magnetic field autocorrelation function, can be expressed by
\begin{equation}
\label{eq:wk_four}
\hat{C}_{\perp}(\vkp)\,=\,\hat{M}_{zz}(\vkp, 0)
\,=\,\frac{1}{2}\hat{w}(\vkp, 0).
\end{equation}
This equation states that the $k_z = 0$ plane of the $M_{zz}$
component (see Eq. (\ref{eq:mcomp})) of the Fourier transformed
magnetic autocorrelation tensor is accessible by the 2-dimensional
Fourier transformation of the RM map. Since all diagonal elements of
the autocorrelation tensor are equal in the case of statistical
isotropy, no more information is required in order to reconstruct the
full magnetic autocorrelation.

The 3-dimensional power spectrum $\hat{w}(k)\,d^3 k$ is related to the
1-dimensional magnetic energy spectrum in the Fourier space
$\varepsilon ^{\obs} _{B} (k)\,dk$ by
\begin{equation}
\label{eq:eb_four}
\varepsilon_B ^{\obs}(k) = \frac{k^2\, \hat{w}(k)}{2(2\pi)^3},
\end{equation}
where $\hat{w} (\vk) = \hat{w} (k)$ because of isotropy.

The combination of Eq. (\ref{eq:crm_four}) and Eq. (\ref{eq:eb_four})
leads to an equation which relates the magnetic energy spectrum
directly to the Fourier transformed RM map 
\begin{equation}
\label{eq:ebrm_four}
\varepsilon_B ^{\obs}(k) = \frac{k^2}{a_1\, A_{\Omega}\, (2\pi)^4} \int
^{2\pi} _0 d\phi\, |\hat{\RM}(k_{\perp})|^2,
\end{equation}
which is equivalent to averaging over rings in k-space. Therefore
integration over $k$ yields the average magnetic energy
density $\varepsilon_B ^{\obs} = \int _{0} ^{\infty} dk \, \varepsilon_B
^{\obs}(k)$ and thus, leads to a magnetic field strength estimate.

Other quantities like the correlation length scales, $\lambda_B$ and
$\lambda_{\RM}$, can also be calculated via integrating over the
correlation function:
\begin{eqnarray}
\label{eq:lrm_four}
\lambda_{\RM} &=& 2 \frac {\int ^{\infty} _0 dk \,\hat{w}(k)}{\int
^{\infty} _0 dk \,k\, \hat{w}(k)} \\
\label{eq:lb_four}
\lambda_B &=& \pi \frac {\int ^{\infty} _0 dk \,k\, \hat{w}(k)}{\int
^{\infty} _0 dk \,k^2\, \hat{w}(k)}.
\end{eqnarray}

From these two equations, it is obvious that the RM autocorrelation
length $\lambda _{\RM}$ has a much larger weight on large-scale
fluctuations than the magnetic correlation length $\lambda
_B$. Therefore one expects that $\lambda _{\RM} > \lambda _B$ in the
natural case of a broadly peaked magnetic power spectrum.

The autocorrelation function and the power spectrum are related by a
Fourier transform as stated by the WKT, therefore one can derive
a relation for the magnetic autocorrelation function $w(r)$ in terms
of the power spectrum $\hat{w}(k)$
\begin{equation}
\label{eq:wr_four}
w(r) = \frac {4\pi}{(2\pi)^3} \int ^{\infty} _{0} dk \,k^2\,
\hat{w}(k) \, \frac {\sin {kr}}{kr}.
\end{equation}
The derivation of the function $w(r)$ employing the power spectra
$\hat{w}(k)$ can be used as a good test for proper numerics, since it can be
compared to the deprojected magnetic autocorrelation function $w(r)$
determined in real space.

\subsection{Testing the model and assessing the influence of the window
function \label{sec:testmodel}}

There are different methods to assess the results obtained using the
approaches in real and Fourier space described above. Besides the
comparison of the numbers determined for the characteristic length
scales and magnetic field strengths in both spaces, there is also the
possibility to evaluate the quality of the models used for the
magnetic energy density and the location of the radio emitting source
within or behind the Faraday screen. For this purpose a $\chi ^2$-test
function can be introduced such that
\begin{equation}
\label{eq:chi}
\chi^2(\vx_{\perp}) = \frac{\RM (\vxp)^2}{\langle \RM(\vxp)^2 \rangle},
\end{equation}
where the expected RM dispersion is defined as $\langle \RM(\vxp)^2
\rangle = 0.5\,\,a_0 ^2 n_{e0}^2 B_0 ^2 \lambda_{B} \int ^{\infty}
_{-\infty} dz f^2(\vx)$. If one averages this function radially and
examines it for any tendency in its spatial distribution, one has a
first test of the model. If there are any large scale trends observed
one should reconsider the model used. Furthermore, the integration of
this function should result in a value close to unity
\begin{equation}
\label{eq:chi_av}
\chi^2 _{av} = \frac {1}{A_{\Omega}} \int dx_{\perp}^2\, \chi ^2 (\vxp)
\approx 1,
\end{equation}
if the choice of the model is a suitable description of the global
magnetic energy distribution and the geometry of the source. Higher
values of $\chi^2 _{av}$ would arise for unsuited models and one would
have to refine the model parameters.

The isotropy which is an important assumption for our analysis can be
assessed by the inspection of the Fourier transformed RM maps. The
assumption of isotropy is valid in the case of a spherically symmetric
distribution of $|\hat {\RM} (k)|^2$.

A completely different problem is the assessment of the influence of
the window function on the shape of the magnetic power spectrum and
thus, on the measured magnetic energy spectrum. As stated above, the
introduction of a finite window function has the effect to smear out
the power spectrum and Eq.(\ref{eq:wk_four}) becomes
\begin{equation}
\label{eq:cexp}
\hat{C}^{\expect} _{\perp} (\vkp) = \frac{1}{2} \int \! d^3q\, \hat{w}(q)\,
\frac{q_{\perp} ^2}{q^2} \, \frac{|\hat{f}(\vkp -
\vec{q})|^2}{(2\pi)^3 V_{[f]}}.
\end{equation}
The term $q^2_{\perp}/q^2 \leq 1 $ in the expression above states some
loss of magnetic power. Without this term, the expression would
describe the redistribution of magnetic power within Fourier
space, where in this case the magnetic energy would be conserved.

However, the expression can be employed as an estimator of the
response of the observation to the magnetic power on a given scale $p$
by setting $\hat{w}(q) = \delta (q-p)$ in the above equation. For an
approximate treatment of a realistic window and a spherical data
average, one can derive
\begin{equation}
\label{eq:wpobs}
\hat{w}^{\expect} _{p} (k_{\perp}) = \frac{2p}{\pi} \int ^{k_{\perp} + q}
_{|k_{\perp} - q|} \!\!\!\!\!\!\!\!\!\!\!\! dq \,\,\,\,\, 
\frac{q \int ^{2\pi} _0 d\phi W_{\perp}(q(\cos{\phi}, \sin{\phi}))}
{\sqrt{4q^2 p^2 - (q^2 + p^2 - k^2 _{\perp})^2}}.
\end{equation} 
Here the projected Fourier window was introduced
\begin{equation}
W_{\perp}(\vkp) = \frac{|\hat{f}_{\perp} (\vkp)|^2}{(2\pi)^2
A_{[f_{\perp}]}},
\end{equation}
where $A_{[f_{\perp}]} = \int d^2 x_{\perp}\,\,f^2 _{\perp} (\vxp)$
while defining the projected window function as $f_{\perp}(\vxp)
= \int dz f^2(\vx)V_{[f]}/A_{\Omega}$.

One can also require for the input power spectrum a field strength by
setting $\hat {w}(q) = 2\pi^2 \, (B ^2/p^2) \, \delta (q-p)$ in
Eq. (\ref{eq:cexp}). The resulting power spectrum can be treated as an
observed power spectrum and thus, a magnetic field strength
$B_{\expect}$ can be derived by integration following
Eq. (\ref{eq:eb_four}). The comparison of $B$ and $B_{\expect}$ for
the different $p$-scales yields a further estimate for reliable ranges
in k-space.

So far the response power spectrum was derived for an underlying
magnetic spectrum located only on scales $p$. One could also add the
particular response functions such that the actual observed power
spectrum $\hat{w}^{\obs}(k)$ would be matched. Typical power spectra,
e.g. for turbulence, show a power law behaviour over several orders of
magnitude. Therefore, one would choose to weight the different
response functions $\hat{w}_{p} ^{\expect} (k_{\perp})$ derived using
Eq. (\ref{eq:wpobs}) by a power law while integrating over them
\begin{equation}
\label{eq:powlaw}
\hat{w}^{\expect}(k) = w_0 \, \int _{p_{\min}} ^{p_{\max}} \!\!\!\!\!\!dp\,
\hat{w}^{\expect} _p (k_{\perp})\,\, \left( \frac{p}{k_0} \right)
^{-\alpha -2},
\end{equation}
where $w_0$ represents a normalisation constant, $k_0$ a typical
$k$-space length scale and $p_{\min}$ and $p_{\max}$ are upper and
lower cutoffs for the integration over the response
functions. Ideally, one would choose the normalisation constant $w_0 =
c_0\,B_0 ^2 /k_0 ^3$, where $c_0$ is chosen such that $\int dk^3 \,
\hat{w}^{\expect}(k) = B_0 ^2$. One would vary the spectral index
$\alpha$ of the power law, $B_0$ or $w_0$ and the lower cutoff of the
integration $p_{\min}$ in order to match the two functions,
$\hat{w}^{\obs}(k)$ and $\hat{w}^{\expect}(k)$.
 
One has direct access to the average magnetic energy density
$\varepsilon_B$ by integrating Eq.(\ref{eq:powlaw}) which results in
the analytic expression
\begin{equation}
\label{eq:ebdirect}
\varepsilon_{B} = \frac{w_0\,k_0 ^3}{2(2\pi)^3\,(\alpha - 1)}\, \left[
\left( \frac{p_{\min}}{k_0} \right) ^{1-\alpha}-\,\,\,\left(
\frac{p_{\max}}{k_0} \right)^{1-\alpha}\right].
\end{equation}

Following the definition of the magnetic autocorrelation length
$\lambda_B$ (Eq. (\ref{eq:lb_four})), one can derive an expression for
this length using the parameters determined in the analysis above
\begin{equation}
\label{eq:k0}
\lambda _B = \pi \frac{\alpha - 1}{\alpha}\, \frac
{p_{\min} ^{-\alpha} - p_{\max} ^{-\alpha}} {p_{\min} ^{1-\alpha} -
p_{\max} ^{1-\alpha}}.
\end{equation}

In the calculation of the expected response $\hat{C}_{\perp}^{\rm
exp}(\vkp)$ and thus, $\hat{w}^{\rm exp} (\vkp)$ to an underlying
magnetic power spectrum $\hat{w}(k)$ seen through some window $W(\vk)$
as expressed by Eq. (\ref{eq:wpobs}) (or Eqs. 51, 67, 68 of
\citet{2003A&A...401..835E}), it has not been considered that the mean
RM is subtracted from the observed RM maps before our analysis is
applied to them in order to remove contributions from foreground RM
screens (e.g. the galaxy). We take this into account by noting that
the RM value of a pixel at $\vxp$ in the map will be changed only in
the case of $\vxp\in \Omega$, which can be written as
\begin{equation}
\RM(\vxp) \rightarrow \RM^*(\vxp) = \RM(\vxp) - \vec{1}_{\{\vxp \in
 \Omega\}} \int \!\!\frac{d^2x_\perp'}{A_\Omega} \, \RM(\vxp').
\end{equation}

Using the properties of the Fourier transform of a function with
compact support, one can show that
\begin{equation}
\hat{C}_{\RM^*}^{\rm exp}(\vkp) = \hat{C}^{\rm exp}_\RM(\vkp) -
\hat{C}^{\rm exp}_\RM(\vec{0}_\perp) |\Delta_\Omega(\vkp)|^2\,,
\end{equation}
where
\begin{equation}
\Delta_\Omega(\vkp) = \frac{1}{A_\Omega} \int \!\!d^2x_\perp
\vec{1}_{\{\vxp \in  \Omega\}}\, e^{i\,\vkp \vec{\cdot}\vxp}\,,
\end{equation}
and $\hat{C}^{\rm exp}_\RM(\vkp)$ is the expected, uncorrected
response to $\hat{w}(k)$. The linearity of the problem ensures that
this correction is valid for any input power spectra $\hat{w}(k)$ (as
long as $\hat{C}^{\rm exp}_\RM(\vkp)$ is calculated from it) which
could be a delta-function, a power law, or could have any other shape.
Similarly, we can write for the spherically averaged expected observed
power spectrum
\begin{equation}
\hat{w}_{*}^{\rm exp}(k) = \hat{w}^{\rm exp}(k) - \hat{w}^{\rm
exp}(0) \frac{1}{2\,\pi} \,\int_0^{2\,\pi} \!\!\!\!\!\! d\phi\,
|\Delta_\Omega(k (\cos\phi,\sin\phi)|^2\,,
\end{equation}
where 
$\hat{w}^{\rm exp}(k)$ is given by Eq. (\ref{eq:wpobs}).
 
The subtraction of the mean RM does not only remove unwanted
homogeneous foregrounds, it also ensures that $\hat{w}_{*}^{\rm
exp}(k) $ (or $\hat{C}_{\RM^*}^{\rm exp}(\vkp)$) vanishes at the
origin $\hat{w}_{*}^{\rm exp}(0) = 0$ (similarly $\hat{C}_{\RM^*}^{\rm
exp}(\vec{0}_\perp) = 0$). For a correct behaviour of the responses at
low $k$ this correction is therefore crucial and will be taken into
account in the following.

\section{Application to existing Rotation Measure Maps of Galaxy
Clusters \label{sec:application}} 

Before we applied our analysis to the data any mean RM value was
subtracted from the RM maps. This is necessary as mentioned above,
since we assume statistically isotropic fields and therefore we have
to remove any RM foreground which would affect our
analysis. Furthermore, very noisy pixels and pixel areas exhibiting
huge differences in the RM values on scales smaller than the beamsize
were removed from the data to decrease the possible influences of
observational artifacts.

The origin of the z-axis was placed in the cluster centre and was used
as reference point $\vx_{\rf}$ for the global distribution of electron
density and magnetic energy. The latter was assumed to scale with the
electron density such that $\langle \vec{B} ^2 (\vx) \rangle ^{1 / 2}
/ \vec{B} _{0} = (n_e(\vx)/n_{e0})^{\alpha_B}$. Unless stated
otherwise, the parameter $\alpha_B$ was chosen to be unity. For
the electron density distribution a standard $\beta$-profile
\footnote {
defined as
\begin{math}
n_{e}(r) = n_{e0}(1+(r/r_c)^2)^{-\frac{3}{2} \beta},
\end{math}
where $r_c$ is called the core radius.
}
was used.

\subsection{3C75, 3C465 \& Hydra A - the data}

The radio source 3C465 (or NGC 7720) is located in the Abell cluster
2634 of richness class I. The redshift of the object is 0.0302
\citep{1999MNRAS.305..259W}. A detailed X-ray study of this cluster
was performed by \citet{1997A&A...327...37S} using ROSAT PSPC
data. They derive for the density profile a core radius of 4.9 arcmin,
a $\beta$ of 0.79 and an electron density at the cluster centre of
$n_{e0} = 0.0012$ cm$^{-3}$. Their analysis revealed indications of a
weak cooling flow in the cluster centre. The scenario of a weak
cooling flow in Abell 2634 is also supported by
\citet{2000MNRAS.312..663W} who analysed ASCA X-ray data.

The radio galaxy 3C75 which is in the centre of Abell 400 also of
richness class I has a redshift of 0.02315
\citep{1991trcb.book.....D}. The X-ray properties of the cluster have
been studied by \citet{2002ApJ...567..716R} using ROSAT PSPC
data. They determine for the gas density profile a core radius of 3.9
arcmin, a $\beta$ of 0.54 and an electron density at the cluster
centre of $n_{e0} = 0.0016$ cm$^{-3}$. There are no indications of a
cooling flow in Abell 400 \citep{2000MNRAS.312..663W}.
    
The details of the radio data reduction of the RM maps for the two
sources above can be found in \citet{2002ApJ...567..202E}. The typical
RM values observed for them ranging between $-250$ rad m$^{-2}$ and
250 rad m$^{-2}$.  The beamsize for the map of 3C75 (Abell 400) is 3
arcsec which translates into 1.4 kpc and for 3C465 (Abell 2634) the
beamsize is 3.8 arcsec which is equivalent to 2.3 kpc for the
cosmology chosen.

Furthermore, it was assumed that the source plane is parallel to the
observer plane since no indication for the Laing-Garrington effect was
found. 

Hydra A (or 3C218) is an extended extragalactic radio source located
at a redshift of 0.0538 \citep{1991trcb.book.....D} in the cluster
Abell 780 of richness class 0. However, we will refer to it hereafter
as Hydra A cluster. Detailed X-ray studies have been performed on it
\citep[e.g.][]{1997ApJ...481..660I, 1998MNRAS.298..416P,
2001ApJ...557..546D}. For the derivation of the electron density profile
parameter, we relied on the work by \citet{1999ApJ...517..627M} done
for ROSAT PSPC data while using the deprojection of X-ray surface
brightness profiles as described in the Appendix A of
\citet{pfrommerensslin}. Since Hydra A is known to exhibit a strong
cooling flow as observed in the X-ray studies, we assumed a double
$\beta$-profile
\footnote {
defined as
\begin{math}
n_{e}(r) = [n_{e1}^2 (0)(1+(r/r_{c1})^2)^{-3\beta}+n_{e2}^2
(0)(1+(r/r_{c2})^2)^{-3\beta}]^{1/2}.
\end{math}
} and used for the inner profile $n_{e1}(0) = 0.056$ cm$^{-3}$ and
$r_{c1} = 0.53$ arcmin, for the outer profile we used $n_{e2}(0) =
0.0063$ cm$^{-3}$ and $r_{c1} = 2.7$ arcmin and we applied a $\beta =
0.77$.  

The details of the radio data reduction can be found in
\citet{1993ApJ...416..554T}. The beamsize of the map is 0.3 arcsec
translating into 0.3 kpc. The source consists of two lobes, a northern
and a southern one. Typical RM values in the north lobe are in the range
between -1000 rad m$^{-2}$ and +3300 rad m$^{-2}$ whereas in the south
lobe also values of down to -12000 rad m$^{-2}$ were observed.

We concentrated our analysis on the north lobe of this source because
the signal-to-noise ratio in the data of the south lobe does not seem
sufficient enough for our purposes. Furthermore, the application of
the data filter to remove big steps between data pixels leads to a
splitting of the RM data in the south lobe into a lot of small
spatially disconnected areas. Such a fragmented window function can
heavily obscure any power spectrum measured.

For Hydra A, there is a clear depolarisation asymmetry of the two
lobes observed as described by the Laing-Garrington effect. Taking
this effect into account and following the analysis of
\citet{1993ApJ...416..554T}, we first rotated the coordinate system
about an angle of 110 degrees such that the x-axis was parallel to the
jet axis. Afterwards the source plane was rotated about the new y-axis
by an inclination angle $\theta$ of 45 degree such that the north lobe
would point towards the observer.

\subsection{Real Space Analysis}
   
\begin{figure}[htb]
\resizebox{\hsize}{!}{\includegraphics{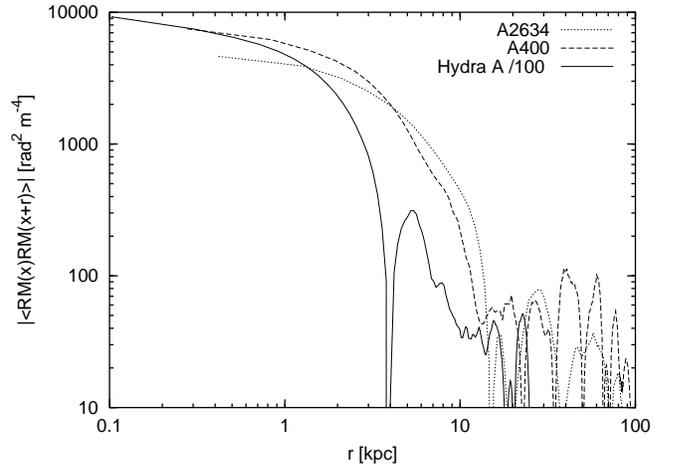}}
\caption[]{\label{fig:correl} RM autocorrelation function
$|C_{\RM}(r_{\perp})|$ of Abell 400, Abell 2634 and Hydra A obtained
using Eq.(\ref{eq:crm_real}). Note that the correlation function for
Hydra A was divided by 100 for representing purposes.}
\end{figure}

After the mean RM subtraction and the noisy pixel filtering, the RM
autocorrelation function $C_{\RM}(r_{\perp})$ was determined using
Eq. (\ref{eq:crm_real}) choosing the described normalisation
scheme. The resulting correlation function for Abell 2634, Abell 400
and Hydra A are exhibited in Fig. \ref{fig:correl}, where the
correlation function of Hydra A was divided by 100 for better
display. The integration of these functions employing
Eq. (\ref{eq:lrm_real}) leads to the information on the RM correlation
length $\lambda_{\RM}$ which was determined to be 7.9 kpc for Abell
2634, 5.3 kpc for Abell 400, and 1.9 kpc for Hydra A.

The determination of the magnetic autocorrelation function in real
space $w(r)$ achieved by the deprojection of the RM autocorrelation is
numerically difficult due to the term in the denominator $\sqrt{y^2 -
r^2}$ of Eq. (\ref{eq:wr1_real}) and Eq. (\ref{eq:wr2_real}) which
represents an integrable singularity and can in principle be avoided
by a well selected coordinate transformation.

\begin{figure*}[htb]
\centering
\includegraphics[width=17cm]{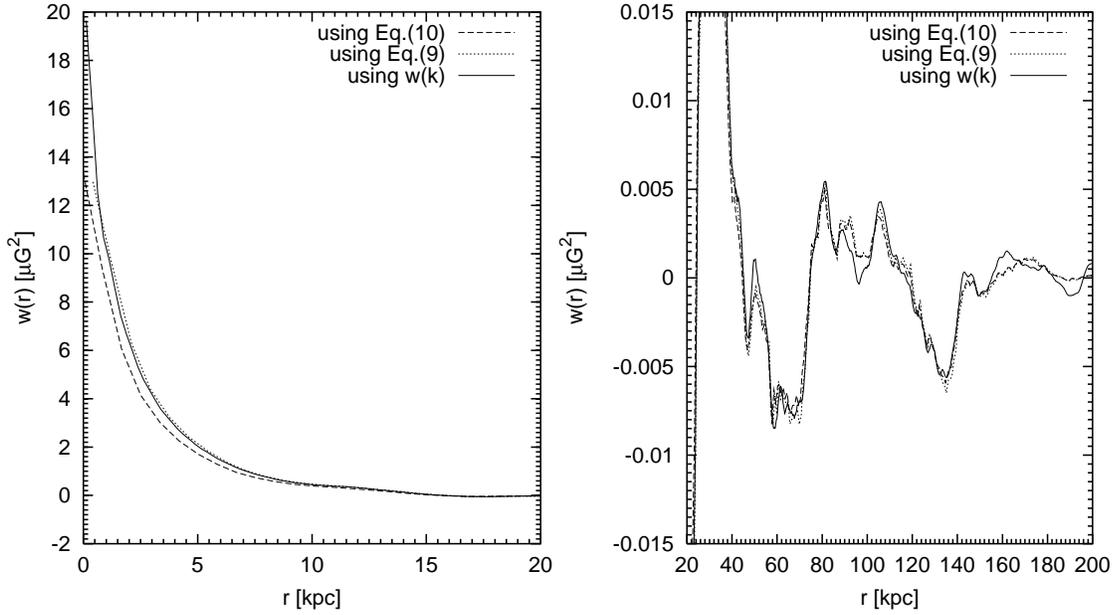}
\caption[]{\label{fig:wr3} The magnetic autocorrelation function
$w(r)$ for Abell 2634 calculated using Eq. (\ref{eq:wr1_real}),
(\ref{eq:wr2_real}) and (\ref{eq:wr_four}). One can clearly see that
the value of $w(r=0)$ can deviate by a factor of two but for larger
$r$ all methods yield almost the same results.}
\end{figure*}

However, the numerical calculation, especially the determination of
the value for $r=0$ of the magnetic autocorrelation function $w(r)$ 
is difficult because an extrapolation to $r = 0$ of the function
$w(r)$ being itself subject to extrapolation and data binning is
involved. Therefore the determination of the point $w(r=0)$ is not
accurate and one should be careful with the interpretation of the
values derived by this method. The behaviour of the magnetic
autocorrelation around $r = 0$ is exhibited in Fig. \ref{fig:wr3} for
the case of Abell 2634, where the magnetic field autocorrelation
functions $w(r)$ derived by employing various methods are shown. One
can see that the steeper the initial slope of the function the less
precise it becomes to determine $w(r = 0)$ since slight deviations in the
slope can lead to very different results.

One should keep this in mind when calculating the value of the
magnetic field autocorrelation length $\lambda_B$ which was determined
to be 6.0 kpc for Abell 2634, 3.9 kpc for Abell 400 and 1.4 kpc for
Hydra A using Eq. (\ref{eq:lb_real}), where for the value of $w(r=0) =
B_{0}^2$ the extrapolation of Eq. (\ref{eq:wr1_real}) to $r = 0$ was
used. Thus, the field strengths were determined to be of about 4
$\mu$G for Abell 2634, of about 9 $\mu$G for Abell 400 and of about 12
$\mu$G for Hydra A in the real space analysis.

\subsection{Fourier Space Analysis}

For the calculations in Fourier space a 2-dimensional Fast Fourier
Transform (FFT) algorithm was employed while setting all blanked
values in the original RM map to zero. The RM autocorrelation function
$\hat{C}_{\perp}(k_{\perp})$ was then obtained by summing over rings
in k-space. Since the magnetic field autocorrelation function
$\hat{w}(k)$ is related to the RM autocorrelation function in Fourier
space simply by multiplying the RM correlation function
$\hat{C}_{\perp}(k_{\perp})$ by two (Eq. (\ref{eq:wk_four})), there is
no numerical difficulty involved deriving this function and thus, the
results for the magnetic field quantities are more precise than those
obtained applying the real space analysis.
 
The integration of the power spectrum $\hat{w}^{\obs}(k)$ following
Eq.(\ref{eq:lrm_four}) and Eq.(\ref{eq:lb_four}) yields the
correlation lengths. Thus, the RM autocorrelation length
$\lambda_{\RM}$ was calculated to be 8.0 kpc for Abell 2634, 5.3 kpc
for Abell 400 and 2.0 for Hydra A. The magnetic field correlation length
$\lambda_{B}$ was determined to be 4.0 kpc for Abell 2634, 2.3 kpc for
Abell 400 and 0.5 kpc for Hydra A in the Fourier analysis.

No numerical problem is involved in the determination of the value of
the magnetic autocorrelation function $w(r)$ for $r = 0$ if one uses
Eq.  (\ref{eq:wr_four}) resulting in magnetic field strengths of about
5 $\mu$G for Abell 2634, of about 11 $\mu$G for Abell 400 and of about
23 $\mu$G for Hydra A.

The application of Eq. (\ref{eq:wr_four}) to the data translates the
magnetic autocorrelation function in Fourier space to real
space. The comparison of the so derived function with the deprojected
functions $w(r)$ in real space is exhibited in Fig. \ref{fig:wr3} for
the case of Abell 2634. One can clearly see that at the origin there
are deviations by a factor of two due to the uncertainties connected
to the extrapolation used in the real space approach. At higher
separations $r$ all three functions shown do not differ significantly
from each other. It is remarkable that except at the origin all three
independent approaches are in such good agreement which demonstrates
reliable numerics.

\begin{figure}[htb]
\resizebox{\hsize}{!}{\includegraphics{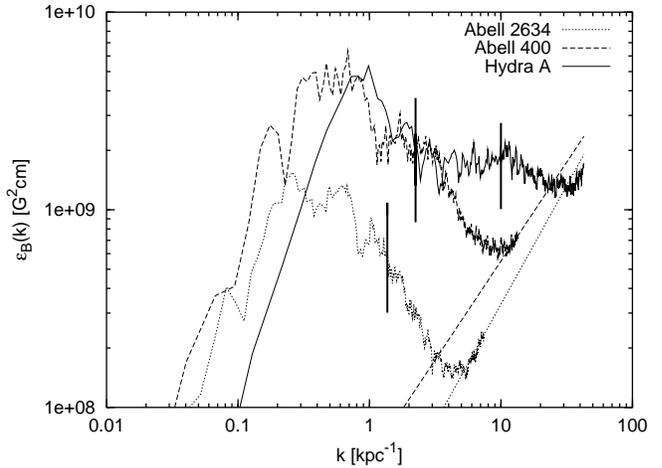}}
\caption[]{\label{fig:power} Magnetic energy spectrum $\varepsilon
_{B} ^{\obs} (k) $ derived for Abell 400, Abell 2634 and Hydra A. The
Fourier transformed beamsizes ($k_{beam} = \pi/l_{beam})$ are
represented by the thick vertical lines. The other straight lines
describe the slope (dashed line represents $\varepsilon_B \propto k$
and the dotted represents $\varepsilon_B \propto k^{1.2}$) of the
increase in the energy density for the largest $k$.}
\end{figure}
 
The knowledge of the 3-dimensional power spectrum $\hat{w}(k)$
also allows calculating the magnetic energy spectrum $\varepsilon _{B}
^{\obs} (k)$ by employing equation (\ref{eq:eb_four}). The results for
the three clusters are shown in Fig. \ref{fig:power}. The magnetic
energy spectra are suppressed at small $k$ by the limited window size
and the subtraction of the mean RM (small $k$ in Fourier space
translate into large $r$ in real space). A response analysis as
suggested above is performed in Sect. \ref{sec:applytest} in order to
understand this influence in more detail.

Another feature of these energy spectra is the increasing energy
density at the largest $k$-vectors in Fourier space and thus, small
$r$ in real space. They can be explained by uncorrelated noise on
scales smaller than the beamsize which possesses a power spectrum of
$\varepsilon_B \propto k$ \citep{2003A&A...401..835E}. The
determination of the increasing slope of the energy densities for
large $k$ revealed proportionalities of $\varepsilon_B \propto
k^{1.0...1.2}$ also shown in Fig. \ref{fig:power}. Thus, we conclude
that at the largest $k$-scales uncorrelated noise dominates the shape
of the energy spectra.

It seems reasonable to introduce an upper cutoff for the integration
of the magnetic energy spectra in Eq. (\ref{eq:eb_four}). In
Fig. \ref{fig:power}, the equivalent beamsize in Fourier space
$k_{beam} = \pi/l_{beam}$ (where $l_{beam}$ is the beamsize in real
space defined as FWHM) is represented by a vertical line for each
cluster which is 1.4 kpc$^{-1}$ for Abell 2634, 2.2 kpc$^{-1}$ for
Abell 400 and 10.0 kpc$^{-1}$ for Hydra A. One can clearly see that
the noise induced increase of the energy density lies on $k$-scales
beyond $k_{beam}$. Therefore, a suitable upper cutoff for any
integration of the magnetic energy spectrum seems to be $k_{beam}$.

\begin{figure}[htb]
\resizebox{\hsize}{!}{\includegraphics{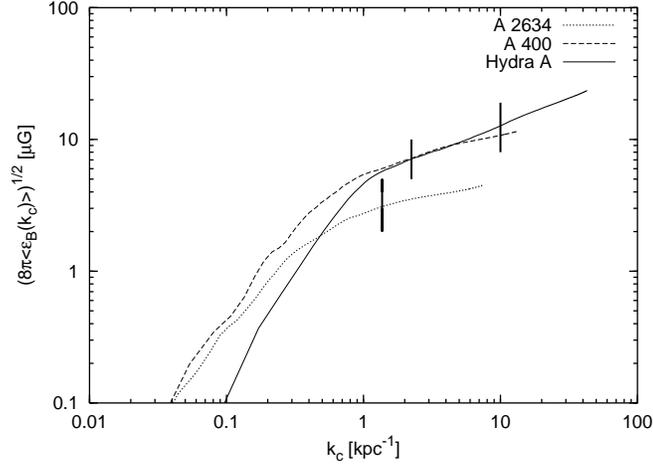}}
\caption[]{\label{fig:cutoff} Magnetic field strength $B$ dependent on
the upper $k$-cutoff in the integration of Eq. (\ref{eq:eb_four}). The
equivalent beamsizes $k_{beam}$ are represented as vertical lines.}
\end{figure}

However, the influence of the upper $k$-cutoff in the integration can
be seen in Fig. \ref{fig:cutoff} which displays the magnetic field
strength $B = \sqrt{8\pi \langle \varepsilon_{B} \rangle}$ estimated
from

\begin{equation}
\varepsilon _{B} (k < k_c) = \int ^{k_c} _0 dk\, \varepsilon_{B} (k).
\end{equation}
 
\subsection{Test the model and assess the influence of the window
function\label{sec:applytest}} 

\begin{figure}[htb]
\resizebox{\hsize}{!}{\includegraphics{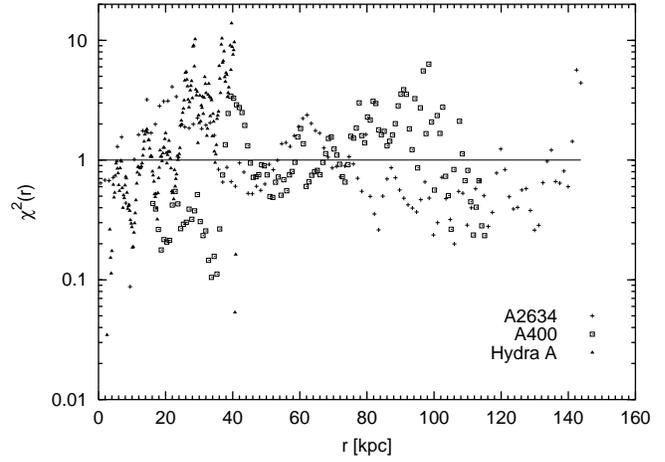}}
\caption[]{\label{fig:chi_squ} Testing the window function by
calculating the $\chi ^2 (x_{\perp})$ distribution for all three
cluster.}
\end{figure}

The various possibilities to test the window function and to assess
its influences on the results outlined in section \ref{sec:model} were
also applied to the data. The $\chi ^2$-function was derived by
radially averaging the $\chi ^2 (\vx_{\perp})$ distribution resulting
from Eq.(\ref{eq:chi}), where for the magnetic autocorrelation length
$\lambda_{B}$ the value derived in the Fourier analysis was used since
this approach seems to be more accurate. The resulting radially binned
distributions are shown in Fig. \ref{fig:chi_squ} determined for the
three clusters. There is no apparent spatial large scale trend visible
for Abell 400 and Abell 2634 which indicates a reasonable model for
the window function. In the case of Hydra A, there appears to be a
trend of higher values for $\chi ^2 (x_{\perp})$ towards larger
$r$, comparable to small scale trends seen in the $\chi ^2
(x_{\perp})$ distribution of Abell 2634 and Abell 400.

Furthermore, the $\chi ^2 (\vxp)$ distribution was integrated
following Eq.(\ref{eq:chi_av}) and $\chi ^2 _{av}$ was calculated to
be 1.0 for Abell 2634, 1.2 for Abell 400 and 1.6 for Hydra A. A
refinement of the model describing parameter does not seem to be
required for Abell 2634 and Abell 400.

Since the value for $\chi ^2 _{av}$ of 1.6 for Hydra A is deviating
from 1, we varied the geometry of the source, i.e. we varied the
inclination angle $\theta$ between source and observer plane from 30
degree up to 60 degree resulting in field strength ranging from 17
$\mu$G to 33 $\mu$G and values for $\chi ^2 _{av}$ of 1.4 to 1.8,
respectively. These values were derived by integrating over the full
accessible unrestricted (without cutoffs) $k$-space.
 
Furthermore, we varied the scaling parameter $\alpha _{B}$ in the
scaling relation of the electron density to the magnetic energy
density. A value for $\alpha _{B}$ of 0.5 is still in the limit of
reasonable values as suggested by
\citet{2001A&A...378..777D}. Furthermore, such a scaling parameter of
$\alpha _{B} = 0.5$ means that the magnetic energy $\varepsilon _B(x)
\propto n_e (\vx)$ and thus, the magnetic field would be proportional
to the thermal energy of the cluster gas assuming approximately
isothermal conditions. However, in this case ($\alpha_B = 0.5$) one
obtains for $\chi ^2 _{av}$ a value of 1.2 and the magnetic field
strength is reduced to 17 $\mu$G by integrating over the full
accessible $k$-space.

One can not be sure that the cause of the trend in the $\chi ^2
(x_{\perp})$ distribution is due to a geometry other than assumed
because it could also be explained by a fluctuation similar to the
one seen in the $\chi ^2 (x_{\perp})$ distributions of Abell 2634 and
Abell 400. There is no reason to change the initial assumption for the
geometry of Hydra A. However, we should keep in mind that the central
field strength given could be slightly overestimated for Hydra A but
it is not clear to what extent if at all.

\begin{figure}[htb]
\resizebox{\hsize}{!}{\includegraphics{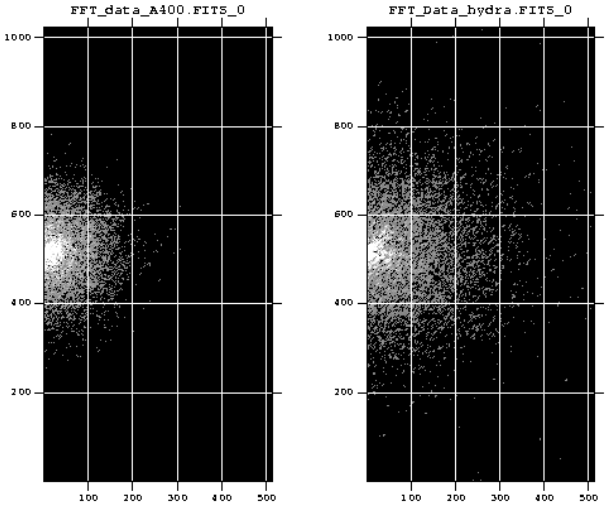}}
\caption[]{\label{fig:fft} The power $|\langle \hat{\RM}(k) \rangle|^2$
of the Fourier transformed RM map of Abell 400 on the left side and
Hydra A on the right side. The image for Abell 2634 looks
similar. Only half of the Fourier plane is exhibited the other half is
point symmetric to the one shown}
\end{figure}

A good test for the validity of the isotropy assumption is the
inspection of the Fourier transformed RM data $|\hat{\RM}(k)|^2$ as
shown in Fig. \ref{fig:fft} for Abell 400 and Hydra A. The FFT
data look similar for Abell 2634. In this figure half of the Fourier
plane is shown since the other half is symmetric to the one
exhibited. No apparent anisotropy is present in this figure.

\begin{figure}[htb]
\resizebox{\hsize}{!}{\includegraphics{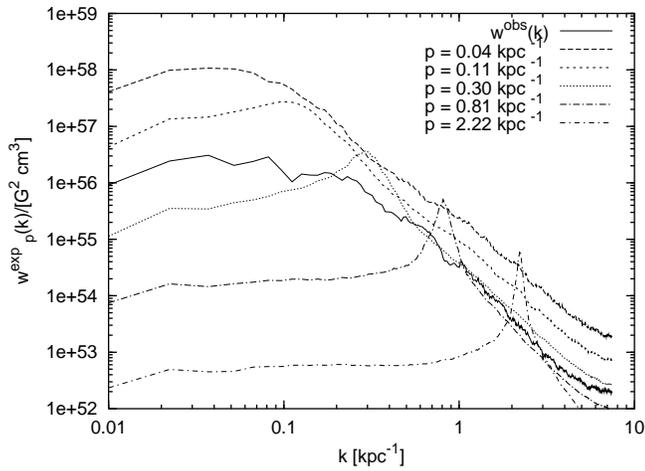}}
\caption[]{\label{fig:response} Responses at various scales $p$ to the
window function of Abell 2634 assuming a $B = 5\mu$G in comparison
to the observed magnetic autocorrelation function $\hat{w}(k)$.}
\end{figure}

We mentioned already that the magnetic energy spectrum $\varepsilon _B
^{\obs}(k)$ in Fig. \ref{fig:power} is suppressed by the limited
window size for small $k$-vectors in Fourier space. Therefore a
response analysis is necessary in order to understand the influence of
the window on the shape of the magnetic energy spectrum. For this
purpose, the projected window function was Fourier transformed by
employing a FFT algorithm and inserted into Eq. (\ref{eq:wpobs}). We
then compared the response functions $\hat{w}^{\expect}(k)$ obtained
to the observed power spectrum $\hat{w}^{\obs}(k)$ as shown in
Fig. \ref{fig:response} for the case of Abell 2634, where we chose
for the normalisation a magnetic field strength $B$ of 5 $\mu$G. The
corresponding figures look very similar for Abell 400 and Hydra A.

From Fig. \ref{fig:response}, one can clearly see that the response to
delta function like input power spectra on small $p$ scales in Fourier
space (i.e. large $r$ in real space) is a smeared out function as one
would expect. The response for larger $p$ becomes more strongly peaked
suggesting that at $k$-scales larger than 0.3 kpc$^{-1}$ for Abell
2634 the influence of the window function becomes negligible. From
similar plots for the other two clusters under consideration, scales
of about 0.4 kpc$^{-1}$ for Abell 400 and 1.0 kpc$^{-1}$ for Hydra A
are found. Thus, one would use this value as a lower cutoff $k_{\min}$
for the determination of the magnetic field strength.

\begin{figure}[htb]
\resizebox{\hsize}{!}{\includegraphics{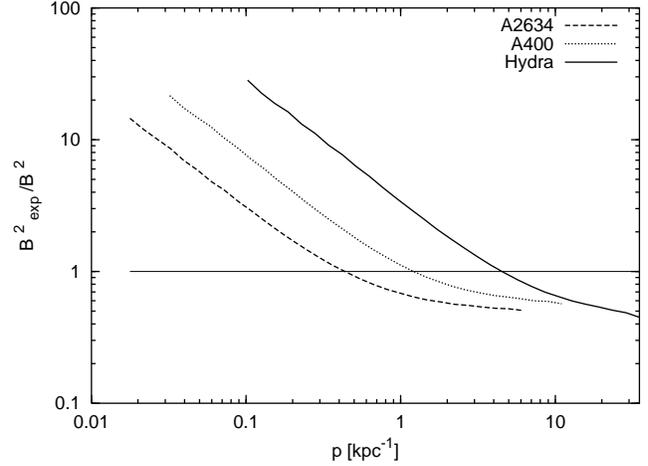}}
\caption[]{\label{fig:bcomp} Comparison between the expected magnetic
field strength $B$ to the measured one $B_{\expect}$ for the
particular $p$-scales for the three clusters.}
\end{figure}

From the response power spectra $\hat {w}_p (k_{\perp})$ calculated
for different $p$ requiring a magnetic field strength $B$, one can
derive a magnetic field strength $B_{\expect}$ by integration. In
Fig. \ref{fig:bcomp}, the comparison between these two field strengths
is plotted for the different $p$-scales. It can be seen that for the
smallest $p$'s the deviation between expected field strength and
actual observed one is significantly but for larger $p$ they are
almost equal suggesting that on these scales the influence of the
window is negligible. Taking Fig. \ref{fig:bcomp} into account, the
values determined above as lower cutoff $k_{\min}$ can be confirmed.

\begin{figure}[htb]
\resizebox{\hsize}{!}{\includegraphics{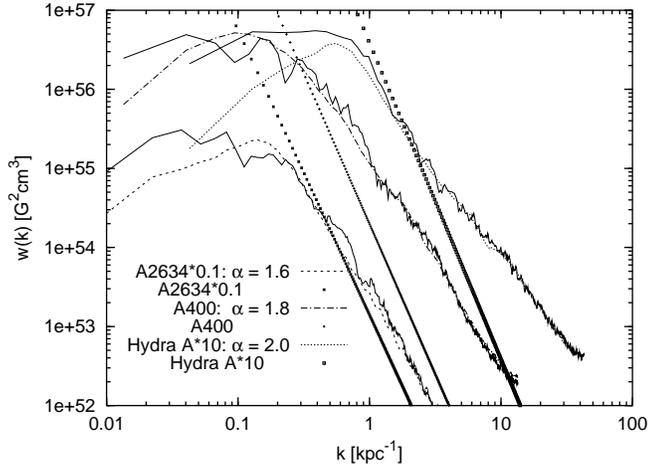}}
\caption[]{\label{fig:sump} Closely $p$-spaced response functions were
integrated by employing Eq. (\ref{eq:powlaw}) and the resulting
$\hat{w}^{\expect}(k) $are exhibited in comparison to the observed
power spectrum $\hat{w}^{\obs}(k)$ shown as solid lines. The spectral
index $\alpha$ used for Abell 2634 is 1.6, for Abell 400 is 1.8 and
for Hydra A is 2.0. The straight lines represent the power law
assuming the spectral index determined from the response
analysis. Note that the data for Abell 2634 are multiplied by 0.1 and
the data for Hydra by 10 for representing purposes.}
\end{figure}

Another possibility to assess the influence of the window on the power
spectra is to match the added response functions and the actual
observed power spectra $\hat{w}^{\obs}(k)$. This was done by applying
Eq. (\ref{eq:powlaw}) to a set of response functions generated for
closely spaced $p$-scales. The resulting function
$\hat{w}^{\expect}(k)$ was matched to the actual observed power
spectrum $\hat{w}^{\obs}(k)$ for each of the three clusters by varying
$B_0$, the spectral index $\alpha$ of the power law in
Eq. (\ref{eq:powlaw}) to match the slope of the functions and
$p_{\min}$ as lower cutoff to fit the function at its turnover for
small $k$-scales. The resulting functions are shown in
Fig. \ref{fig:sump} in comparison to the respective observed power
spectra. The shape of the power spectra was matched for Abell 2634 for
a spectral index of $\alpha = 1.6$, for Abell 400 for an $\alpha =
1.8$ and for Hydra A for an $\alpha = 2.0$ whereas a lower $k$-cutoff
of 0.09 kpc$^{-1}$ was used for Abell 2634, 0.08 kpc$^{-1}$ for Abell
400 and 0.3 kpc$^{-1}$ for Hydra A.

The spectral index $\alpha$ can also be used to plot the respective
power law as represented by the straight lines in
Fig. \ref{fig:sump}. The slopes of these lines clearly deviate from
the respective observed power spectra suggesting that one cannot
estimate differential parameters like spectral indeces directly from
Fourier transformed RM maps. More sophisticated approaches should be
developed. However, our approach allows to exclude flatter spectral
slopes than $\alpha = 1.3$ still leaving a Kolmogorov spectrum
($\alpha = 5/3$) as a possible description. For attempts to measure
the magnetic power spectrum from cluster simulations and radio maps
see \cite{2002A&A...387..383D} and \cite{astro-ph/0211292}
respectively\footnote{The conventions describing the spectra may
differ in these articles from the one used here.}.

\section{Results\label{sec:discussion}}

A summary of the values for RM autocorrelation length $\lambda_{\RM}$,
magnetic field autocorrelation length $\lambda_B$ and the central
magnetic field strength $B_0$ for the magnetised cluster gas in Abell
2634, Abell 400 and Hydra A, derived in Fourier space and real
space is given in Tab. \ref{tab:res}. The RM autocorrelation length
$\lambda_{\RM}$ determined in both spaces is almost equal which
demonstrates reliable numerics. Furthermore, it is found that the RM
autocorrelation length $\lambda_{\RM}$ is larger than the magnetic
autocorrelation length $\lambda_B$. The deviation of the magnetic
field quantities between both spaces is caused by the numerically
complicated deprojection of the magnetic autocorrelation function
$w(r)$ in real space. Thus, the results in Fourier space are more
reliable and should be considered for any interpretation and
discussion.
 
In Sect. \ref{sec:applytest}, we discussed many possible influences on
the power spectra. We verified that for intermediate $k$-ranges
extending over at least one order of magnitude the power spectra are
not governed by the window or the resolution of the RM maps and thus,
represent most likely the magnetic field properties of the cluster
gas. Thus, one can derive field quantities such as magnetic field
strength $B_0$ and autocorrelation length $\lambda_B$ from this
intermediate $k$-range.

The suitable cutoffs for the integration of the power spectra in order
to derive the magnetic field strength and autocorrelation length were
determined in Sec. \ref{sec:applytest}. Using for Abell 2634 a lower
cutoff $k_{min}$ of 0.3 kpc$^{-1}$ and an upper cutoff of 1.4
kpc$^{-1}$ results in a magnetic field strength of 3
$\mu$G. Integrating the power spectrum of Abell 400 in the range from
0.4 kpc$^{-1}$ to 2.2 kpc$^{-1}$ yields a field strength of 6
$\mu$G. The same was done for the Hydra A cluster in the range from 1
kpc$^{-1}$ to 10 kpc$^{-1}$ which resulted in a field strength of 13
$\mu$G. The determination of the magnetic field autocorrelation length
$\lambda_B$ by integration over the power spectra in limited $k$-space
yields 4.9 kpc for Abell 2634, 3.6 kpc for Abell 400 and 0.9 kpc for
Hydra A. 

In the course of the response analysis in order to estimate the
influence of the window on the power spectra, we matched the
integrated response power spectra $\hat {w}^{\expect} (k)$ with the
actual 3-dimensional observed power spectra $\hat {w} ^{\obs}(k)$ and
determined the spectral index $\alpha$, $c_0$ and the lower cutoff
$p_{\min}$ in Eq. (\ref{eq:powlaw}). Using these values for the
parameters in Eq. (\ref{eq:ebdirect}), the direct integration in the
limits of $p_{\min}$ and $p_{\max} = k_{\rm {beam}}$ of the power
spectra obtained by the response analysis results in magnetic field
strength of about 3 $\mu$G for Abell 2634, of about 6 $\mu$G for Abell
400 and of about 11 $\mu$G for Hydra A. These field strengths are in
good agreement with the results obtained by the restricted Fourier space
integration of the observed 3-dimensional power spectra. 

However, the values for $\lambda_B$ were determined for the same
$p_{\min}$ and $p_{\max} = k_{\rm {beam}}$ applying
Eq. (\ref{eq:k0}). The resulting value is 13 kpc for Abell 2634, 17
kpc for Abell 400 and 4 kpc for Hydra A. They deviate by a factor of 4
to 5 from the magnetic field autocorrelation length $\lambda_B$
determined from the restricted Fourier space integration. The reason
for this behaviour could be found in the sensitivity to the lower
cutoff $p_{\min}$ and the power law index $\alpha$. We have already
mentioned that our method might not be suitable at that stage to
determine differential parameters such as a power law index $\alpha$. A
more detailed discussion which is beyond the scope of this paper is
necessary in order to understand this behaviour.

However, the values of the magnetic field strength $B_0$ for Abell 400
and Abell 2634 of about 6 $\mu$G and 3 $\mu$G, respectively, are larger
compared to the initial analysis performed by
\citet{2002ApJ...567..202E} for the case of a magnetic sheet with a
thickness of 10-20 kpc. They discuss a variety of magnetic field
models and we refer to Tab. 2 of their paper for detailed numbers and
description.

\begin{table*}[htb]
\begin{center}
\begin{tabular}{ l c c c c c c c c c} 
\hline & & \multicolumn {3}{c}{Real Space} &
 \multicolumn{3}{c}{non-restricted Fourier Space} &
 \multicolumn{2}{c}{restricted Fourier Space} \\ 
cluster & $\alpha_{B}$ &
 $\lambda_{\RM}$ [kpc] & $\lambda_{B}$ [kpc] & B$_{0}\,[\mu$G] &
 $\lambda_{\RM}$ [kpc] & $\lambda_{B}$ [kpc] & B$_{0}\,[\mu$G] &
 $\lambda_{B}$ [kpc] & B$_{0}\,[\mu$G] \\ 
\hline
 Abell 400  & 1.0 & 5.2  & 3.9 & 8  & 5.3 & 2.3 & 12 & 3.6 & 6 \\
 Abell 2634 & 1.0 & 7.9  & 6.0 & 4  & 8.0 & 4.0 & 5 & 4.9 & 3 \\
 Hydra A    & 1.0 & 1.9  & 1.4 & 13 & 2.0 & 0.5 & 23 & 0.9 & 13 \\
\hline
 Abell 400  & 0.5 & 5.2  & 3.9 & 7  & 5.3 & 2.3 & 10 & 3.4 & 6 \\
 Abell 2634 & 0.5 & 7.9  & 6.0 & 3  & 8.0 & 4.0 & 4 & 4.9 & 3 \\
 Hydra A    & 0.5 & 1.9  & 1.4 & 10 & 2.0 & 0.5 & 17 & 0.9 & 9 \\

\end{tabular}
\end{center}
\caption{Values for the autocorrelation length scales $\lambda_B$ and
$\lambda_{\RM}$ and the magnetic field strength $B_0$ at the cluster
centre obtained in real space analysis, in Fourier space analysis and
in restricted Fourier Space analysis are given for the different
clusters under consideration. The values are calculated for two
different scaling parameters $\alpha_B$. Note that for position other
than the cluster centre the average magnetic energy is given by
$\langle \vec{B}^2(\vx) \rangle ^{1/2} = \vec{B}_0\,
(n_e(\vx)/n_{e0})^{\alpha_B}$.}
\label{tab:res}
\end{table*}

For the Hydra A cluster, \citet{1993ApJ...416..554T} conclude in their
analysis that there is a random magnetic field component of about 30
$\mu$G with a correlation length of 4 kpc. The deviation by a factor
of 3 of the field strength from the value revealed by our analysis
might be explained by the usage of an improved electron density
profile for our analysis which also takes the cooling flow into
account. Our value for the central magnetic field strength $B_0$ might
be also lower due to the conservative approach of restricting the
$k$-space integration range. Another explanation for the difference
could be sought in our exclusion of the south lobe from the
calculation above. Including the south lobe in our analysis
leads to higher central field strength but given the influence of the
very complicated window function in the case of the south lobe, it is
not clear to what extent the real power spectrum is resembled.
 
The estimation of the dynamical importance of the magnetic fields
derived for the cluster gas can be done by comparing the thermal
pressure ($p_{th} = 2n_e (0)kT_{core}$) with the magnetic pressure
($p_{B} = B_0 ^2 / (8 \pi)$). One can calculate $p_{B}/p_{th}$ which
yields 0.08 for the case of Abell 2634 (assuming a $T_{core} = 1.2$
keV \citep{1997A&A...327...37S}), $p_{B}/p_{th}$ = 0.19 for Abell 400
($T_{core} = 1.5$ keV \citep{2002ApJ...567..202E}) and $p_{B}/p_{th}$
= 0.01 for Hydra A ($T_{core} = 2.7$ keV
\citep{2001ApJ...557..546D}). It is astonishing that the value of
$p_{B}/p_{th}$ is smaller for Hydra A, which is a cooling flow
cluster, than for the non-cooling flow cluster Abell 400 and Abell
2634. The values of $p_{B}/p_{th}=0.1...0.2$ for the latter two
clusters give an indication that for those clusters the magnetic field
is of some weak dynamical importance for the cluster gas.

\section{Conclusions\label{sec:conclusion}}
We presented the application of a new analysis of Faraday rotation
maps recently developed by \citet{2003A&A...401..835E} to
observational data in order to estimate magnetic field strength and
autocorrelation length and in order to determine the magnetic power
spectra of the magnetised intra-cluster gas. We described that the
analysis relies on the assumption that the magnetic fields are
statistically isotropically distributed throughout the Faraday
screen. We introduced the window function through which any virtually
statistically homogeneous magnetic field can be thought to be
observed. This window function describes the geometry of the source
and the global properties of the intra-cluster gas such as the
electron density, known from X-ray measurements, and the global
average magnetic field distribution, which we assume to scale with the
electron density. Furthermore, we explained two possible approaches in
real and Fourier space and outlined the tests for the evaluation of
any influence especially arising from the observational nature of the
data such as limited source size, resolution and pixel noise on the
results obtained. However, we stated that our analysis allows to
measure average magnetic energies but it is not sensitive to the
particular realisation of the magnetic field structure.

We applied this approach for the first time to observational data and
derived not only reliable results for magnetic field strength and
autocorrelation length but also we got insight into strength and
quality of any results obtained by applying this new analysis to any
observational data. In order to understand the possible impacts on the
results we reanalysed the Faraday rotation maps of three extragalactic
extended radio sources, i.e. 3C75 in the cluster Abell 400, 3C465 in
Abell 2634, which were kindly provided by Frazer Owen and Jean Eilek,
and Hydra A in Abell 780, which was kindly provided by Greg Taylor. We
performed the analysis in real and Fourier space for these three
Faraday rotation maps. While discussing the difficulties involved in
the application to the data, we realised that the calculations in
Fourier space are more reliable.

We tested the isotropy assumption and no indication of anisotropy was
found. Furthermore a $\chi^2$-test was performed in order to assess
the model adopted for the geometry of the sampling volume
incorporating the global electron density and average magnetic energy
distribution. In the case of Abell 2634 and Abell 400, no indication
was found that our model may be incorrect. However, in the case of
Hydra A we found indications that the window function needs refinement
but the indications were not strong enough in order to enforce this
refinement.

We realised that the magnetic energy spectra $\varepsilon_B(k)$ of the
three clusters investigated are dominated on the largest $k$-scales
(i.e. smallest $r$-scales) by uncorrelated noise. Therefore it seemed
natural to introduce a higher $k$-cutoff $k_{\max}$ for any
integration in $k$-space necessary to derive magnetic field quantities
such as field strength and autocorrelation length. Being conservative,
we used as higher $k$-cutoff the equivalent beamsize in Fourier space
$k_{\max} = \pi/l_{beam}$ which is equal to 1.4 kpc$^{-1}$ for Abell
2634, 2.2 kpc$^{-1}$ for Abell 400 and 10 kpc$^{-1}$ for Hydra A. This
will have the effect of loosing some power which is redistributed due
to the window from smaller to these larger $k$-scales.

On the smallest $k$-scales, i.e. the largest $r$-scales, power is
suppressed because of the limited size of the window. For the
assessment of the influence of this suppression on the power spectra,
we applied a response analysis described in Sec. \ref{sec:testmodel}
to the observational windows. The response of the window to delta like
input power spectra on small $p$-scales is a wide, smeared out
function whereas the response on larger $p$-scales is a peaked
function. This observation motivated the introduction of a lower
$k$-cutoff $k_{\min}$ in any integration in $k$-space. This value was
determined to be 0.3 kpc$^{-1}$ for Abell 2634, 0.4 kpc$^{-1}$ for Abell
400 and 1.0 kpc$^{-1}$ for Hydra A. However, the magnetic field
strengths derived are not sensitive to this lower cutoff due to the
small $k$ power suppression.  

We verified that the intermediate $k$-ranges between $k_{\min}$ and
$k_{\max}$ extending over at least one order of magnitude can be used
to determine actual magnetic field properties of the intra-cluster gas.

Integrating over the response power spectra on particular $p$-scales
enables us to match the so determined power spectra and the actually
observed power spectra in order to have an independent estimate for
magnetic field strengths and spectral slopes. For the three clusters
under consideration, spectral indices $\alpha$ for the slopes of the
power spectra were determined to be in the range from 1.6 to
2.0. Therefore it would be possible that the magnetic field in these
clusters have a Kolmogorov power spectrum exhibiting $\alpha$ =
5/3. However, presently we can not exclude steeper spectra but flatter
spectra exhibiting slopes smaller than $\alpha = 1.3 $ would have been
recognised by our analysis. Although we realised that this analysis is
not suitable for the determination of differential parameters such as
the spectral slopes of power spectra directly from Fourier transformed
RM maps, the determination of integrated quantities such as the
magnetic field strength appears feasible.

Taking all these arguments into account we determined by integration
of the magnetic energy spectrum in the limits between $k_{\min}$ and
$k_{\max}$ values for the RM autocorrelation length $\lambda_{\RM}$,
the magnetic field autocorrelation length $\lambda_B$ and the central
average magnetic field strength $B_0$. An overview of these numbers is
given in Tab. \ref{tab:res} where these values are compared to those
derived in real space and non-restricted Fourier space.

The magnetic field autocorrelation length $\lambda_B$ was determined
for the restricted Fourier space integration to be 4.9 kpc for Abell
2634, 3.6 kpc for Abell 400 and 0.9 kpc for Hydra A. In comparison
with the RM autocorrelation length $\lambda_{\RM}$  calculated to be
8.0 kpc for Abell 2634, 5.3 kpc for Abell 400 and 2.0 kpc for Hydra A,
it can be said that these two characteristic length scales differ from
those often assumed in previous work.

The magnetic field strength in the cluster centre $B_0$ was calculated
for the same limited $k$-space and was determined to be 3 $\mu$G for
Abell 2634, 6 $\mu$G for Abell 400 and 13 $\mu$G for Hydra A. Given
the assumption of isotropy and a scaling parameter $\alpha_B = 1.0$,
these are conservative values. The resulting magnetic pressures
suggest some small but non negligible pressure support for a dynamical
influence in the case of Abell 2634 and Abell 400
since the value of $p_{B}/p_{th}$ is 0.08 and 0.19, respectively.

We note, that our approach so far is not able to separate the
influence of noise on the maps from the astrophysically interesting
signal of intra-cluster magnetic fields. Investigations aiming to
detect and minimise such spurious signals are in preparation
(\citet{2003ApJ...Ensslin}; Dolag et al., in prep.).

\begin{acknowledgements}
We thank Jean Eilek, Frazer Owen and Greg Taylor who kindly provided
us the Faraday rotation maps of 3C75, 3C465 and Hydra A. We
acknowledge lively discussions with Greg Taylor, Frazer Owen, Jean
Eilek, Tracy Clarke and Christoph Pfrommer. We thank Matthias
Bartelmann for comments on the manuscript. C.V.\ likes to thank for
the hospitality at the National Radio Astronomy Observatory (NRAO) in
Soccorro in summer 2002, where some of this work was done and lots of
interesting discussions took place. C.V.\ would like to thank the
Deutscher Akademischer Austauschdienst (DAAD) for granting a 'DAAD
Doktorandenstipendium' during the time of the stay at NRAO. This
research has made use of the NASA/IPAC Extragalactic Database (NED)
which is operated by the Jet Propulsion Laboratory, California
Institute of Technology, under contract with the National Aeronautics
and Space Administration. This work was done in the framework of the
EC Research and Training Network \emph{The Physics of the
Intergalactic medium}.

\end{acknowledgements}


\bibliographystyle{aa}
\bibliography{cori}

\end{document}